\documentclass[aps,prb,showpacs]{revtex4}
\usepackage{graphicx}

\begin{document} 
%\draft 
%\preprint{Preprint }
\title{ Charge transport and shot noise in ballistic graphene sheet}  
\author{ E.B. Sonin}

\affiliation{ Racah Institute of Physics, Hebrew University of
Jerusalem, Jerusalem 91904, Israel\\
and \\
Low Temperature Laboratory, 
Helsinki University of Technology, FIN-02015 HUT, Finland} 

\date{\today} 

\begin{abstract}
The current and the shot noise  in a graphene sheet were analyzed in the ballistic regime for arbitrary voltage drops between leads and the sheet in the limit of infinite aspect ratio of the sheet width  $W$ to its length $L$, when quantization of transversal wave vectors is not essential. The cases  of coherent and incoherent ballistic transport were compared. At high voltages  the difference with coherent transport is not essential. But at low voltages conductance and Fano-factor dependences for incoherent transport become non-monotonous so that the conductance has a minimum and the Fano factor has a maximum at non-zero voltage bias.
\end{abstract} 
\pacs{73.23.Ad,73.50.Td,73.63.-b}
\maketitle

\begin{figure}[t]
\centerline{\includegraphics[width=0.8\linewidth]{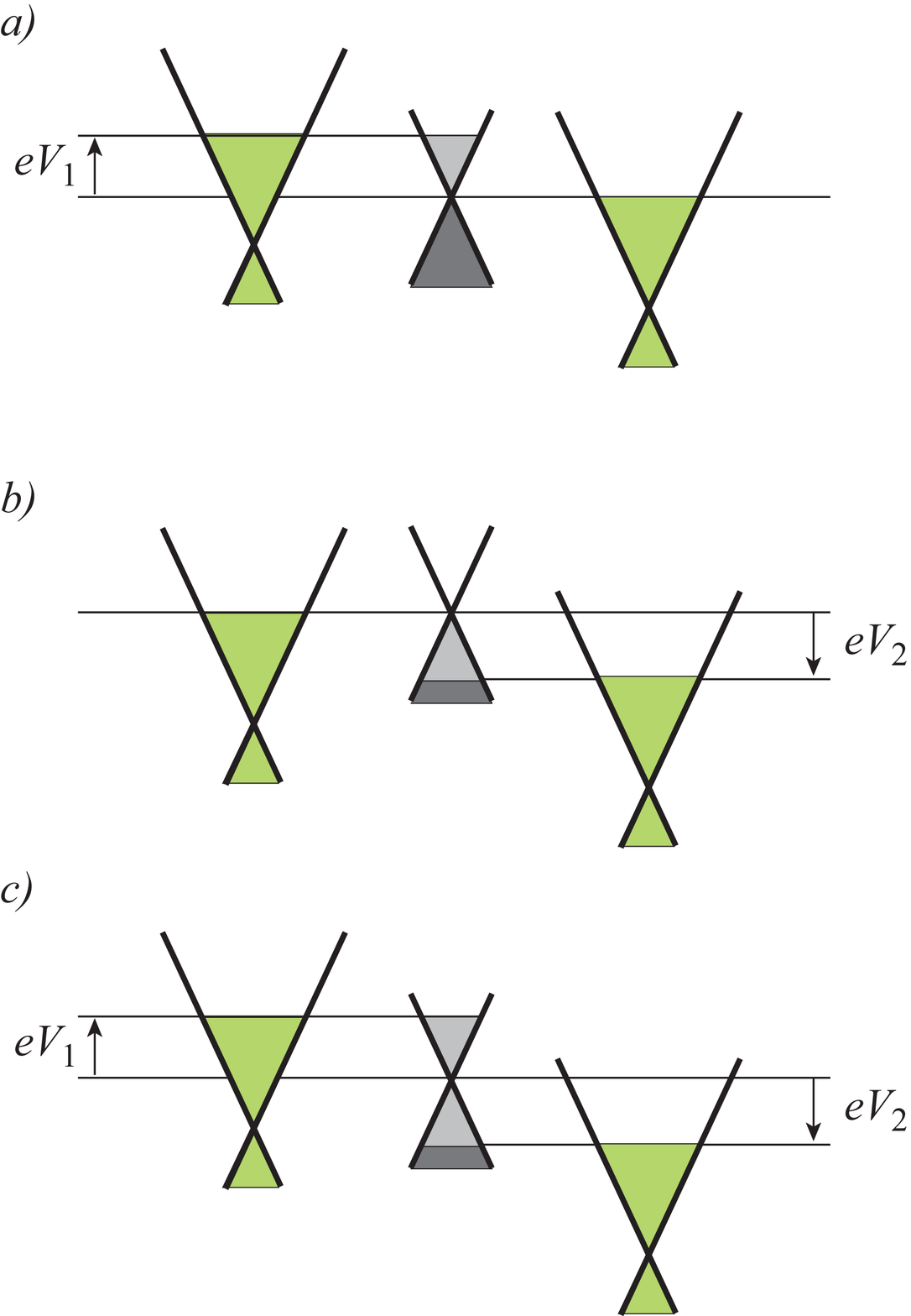}}
\caption{(color online)Energy bands in the left and the right leads and the graphene sheet (center).
a) Voltage bias $V_1>0$ is applied between the left lead and the Dirac point in the sheet, $V_2=0$. b)  
Voltage bias $V_2<0$ is applied between the Dirac point in the sheet and the left lead, $V_1=0$. 
c) Both the voltage drops, $V_1$ and $V_2$, are present. The darker areas in the sheet energy pictures show the fully occupied states in the sheet. The lighter-shaded areas show the partially occupied states (only left-moving particles and right-moving holes). }\label{fig1}
\end{figure}

Transport and shot noise are in the focus of recent experimental and theoretical investigations of graphene since they are tightly connected with unique spectrum of excitations in this   material: the two-dimensional relativistic spectrum  with the Dirac point. The presence of the Dirac point was  experimentally confirmed in graphite (multilayered graphene)\cite{Kop} and graphene\cite{Geim,Kim}. 
From the very first experiments on graphene\cite{Geim,Kim} it was revealed that graphene  has the minimum of conductivity, which is on the order of conductance quantum $e^2/h$ (see also Ref. \onlinecite{Lau}). Katsnelson\cite{Kats} and Tworzyd{\l}o {\em et al.}\cite{Been} have shown theoretically that that the conductivity minimum must appear even in the ballistic regime, i.e., in the absence of disorder. Tworzyd{\l}o {\em et al.}\cite{Been} have also analyzed shot noise and shown that close to the Dirac point  the Fano factor of the graphene sheet coincides  with that of the diffusive wire ($F=1/3$). This stimulated experimental investigations of shot noise\cite{Marc,Hak}.

The analytical results of Katsnelson\cite{Kats} and Tworzyd{\l}o {\em et al.}\cite{Been} were obtained for  the case when the Fermi levels of leads are very close to the Dirac point of the sheet, and the main contribution to current and shot noise came from evanescent states, i.e., from states classically forbidden in the sheet. In the case of the large voltage drops between the leads and the sheet the main contribution to the transport current and the noise power comes from the propagating modes, and the numerical calculations  by Tworzyd{\l}o {\em et al.}\cite{Been}  for the finite aspect ratio $W/L=5$ demonstrated oscillations of the current and the noise power as functions of the large voltages. The question arises, whether these oscillations are connected with the finite aspect ratio, i.e. whether they remain in the limit   $W/L\to \infty $. Also, since the oscillations are related with coherent transport across the graphene sheet, they must be absent if transport is incoherent, i.e., a particle propagating  in the sheet does not keep information on its quantum-mechanical phase. The present Communication focuses on these two issues: current and shot noise at large voltages for the infinite aspect ratio $W/L$, and comparison of coherent and incoherent ballistic transport.

The analysis has been done for the case when leads are also made from graphene but strongly doped so that they have large Fermi wave vectors far from the Dirac point. Katsnelson\cite{Kats} and Tworzyd{\l}o {\em et al.}\cite{Been} considered exactly this case, and we can use their boundary conditions at two contacts (continuity of two-component electron wave function).  Contacts are supposed to be ideal, and the only source of reflection is a mismatch of normal  group velocities  in the lead and in the sheet. Recent detailed theoretical investigations of contacts of graphene with various types of leads \cite{schom,blant} have demonstrated insensitivity of transport through graphene to  detailed properties  of leads.

We consider the graphene sheet of the length $L$ and the width $W$ with the large aspect ratio $W\gg L$. The Fermi levels of the left and the right lead are shifted with respect to the Dirac point in the graphene sheet with the energies $eV_1$ and $eV_2$ respectively, so the voltage bias, which drives the current, is $V=V_1-V_2$ (see Fig.~\ref{fig1}).  The Hamiltonian of the graphene is well known: $H= \hbar v_F (\hat \sigma_x \hat p_x+ \hat \sigma_y \hat p_y)$, where $v_F$ is the Fermi velocity, $ \hat \sigma_x$ and $ \hat \sigma_y$ are the Pauli matrices of the isospin, and $ \hat p_x$ and $ \hat p_y$ are inplane components of the momentum operators. The solutions  of the Dirac equation are the two-component spinors, and the wave function for the electron crossing the left contact from the sheet ($x>0$) to the left lead ($x<0$) is 
\begin{widetext}
 \begin{eqnarray}
\left\{\left( \begin{array}{c} 1  \\ {-k_x+ik_y\over k}\end{array}\right)e^{-ik_x x} 
+r_1 \left( \begin{array}{c} 1  \\ {k_x+ik_y\over k}\end{array}\right)e^{ik_x x}\right\} e^{ik_yy}& \mbox{at}~ x>0,
\nonumber \\
t_1\sqrt{Kk_x\over kK_x}\left( \begin{array}{c} 1  \\ {-K_x+ik_y\over K}\end{array}\right)e^{-iK_x x+ik_yy} 
 & \mbox{at}~ x<0.
              \end{eqnarray}
\end{widetext}
Because of strong doping in the leads the values of the wave vectors in the leads $K\approx K_x$ essentially exceeds the components $k_x$ and $k_y$ of the wave vectors in the sheet. On the other hand, because of the translational invariance along the axis $y$ $K_y =k_y$.
The reflection and transmission amplitudes are determined from the boundary conditions\cite{Been} of continuity of the two spinor components. In the limit $K\gg k$ they are given by
\begin{eqnarray}
r_1=\frac{k _x-k-ik_y }{k_x+k + ik_y}
,~~
t_1= \frac{2 \sqrt{k_xk}}{k_x+k + ik_y},
           \end{eqnarray}
and do not depend on $K$. This is another evidence that ballistic transport in graphene is insensitive to details of leads.

Similarly to this we can find the wave function  for the electron crossing the right contact from the sheet ($x<L$) to the right lead ($x>L$). The reflection and transmission amplitudes in this case are $r_2=r_1^* e^{2ik_x L}$ and $t_2=t_1^* e^{i(k_x-K_x)L}$. 
According to the theory of a double contact \cite{Dat} the total transmission amplitude for phase-coherent transport from the left to the second lead is 
\begin{eqnarray}
t={t_1t_2\over 1-r_1r_2e^{2ik_xL}}
%\nonumber \\
=\frac {4k_xk}{(k+k_x)^2 +k_y^2 -[(k-k_x)^2 +k_y^2]e^{2ik_xL} },
           \end{eqnarray}
whereas the probabilities of   transmission ($T=|t|^2$) and  reflection ($R=|r|^2$) are
\begin{eqnarray}
T=1-R={|t_1t_2|^2\over |1-r_1r_2e^{2ik_xL}|^2}=\frac {k_x^2}{k_x^2 +k_y^2\sin^2(k_xL) }.
           \end{eqnarray}
The analytic continuation of this expression  yields the transmission probability for the evanescent mode $k_x=ip$:
\begin{eqnarray}
T=\frac {p^2}{p^2 +k_y^2\sinh^2(pL) }.
     \label{cor}        \end{eqnarray}
These expression are known from Refs. \onlinecite{Kats} and  \onlinecite{Been}. The charge transfer via evanescent modes is in fact a direct single quantum-mechanical tunneling through the the whole sheet, which is classically inaccessible because of small energy of the electron. 

The total current consists from the current through the upper conductivity band (electrons) and   the current through the lower valence band (holes), i.e. above and below the Dirac point respectively. We consider the case $V=V_1>0,~V_2=0$, when the whole current flows through the upper band. First let us restrict ourselves with propagating modes. Their contributions to  the current and the differential conductance are given by
\begin{widetext}
\begin{eqnarray}
j_p ={2e W\over \pi ^2 } \int_0^{eV/\hbar v_F}dk_x   \int _0^{\sqrt{(eV/\hbar v_F)^2-k_x^2}} dk_y  v_x T
={2ev_F W\over \pi^2} \int_0^{eV/\hbar v_F} dk\int _0^{k} dk_y \frac {k_x^2}{k_x^2 +k_y^2\sin^2(k_xL) }
\nonumber \\
= {ev_F W\over \pi^2} \int_0^{eV/\hbar v_F} {k_x\,dk_x \over  \cos(k_xL)}\ln \frac{V+\sqrt{V^2-(v_F\hbar k_x/e)^2}\cos(k_xL)}{V-\sqrt{V^2-(v_F\hbar k_x/e)^2}\cos(k_xL)},
      \label{prop}     \end{eqnarray}
\begin{eqnarray}
{dj_p\over dV} = G_0L \int_0^{eV/\hbar v_F} {k_x^3\,dk_x \over \sqrt {(eV/\hbar v_F)^2-k_x^2} 
[(eV/\hbar v_F)^2 \sin^2(k_xL) +k_x^2 \cos^2(k_xL) ]}.
     \end{eqnarray}
     \end{widetext} 
Here $G_0=4e^2 v_F W/ \pi hL$ is the minimum conductance,  and $v_x=\partial \varepsilon/\hbar \partial k_x=v_F k_x/k$ is the $x$-component of the group velocity determined by the linear spectrum $\varepsilon=\hbar v_F k$. Because of large aspect ratio $W/L$ we replaced  the sum over transversal components $k_y$ by an integral. In this limit the result does not depend on boundary conditions on lateral sheet edges discussed in Refs. \onlinecite{Kats} and  \onlinecite{Been}.

At high voltages $V\gg \hbar v_F /eL$
\begin{eqnarray}
j_p ={0.785 ev_F W\over \pi}\left({eV\over \hbar v_F}\right)^2.
         \end{eqnarray}

At low voltages the transport via evanescent modes is more important. The expressions for the current and the conductance through evanescent modes are ($p^2=k_y^2-k^2$) 
\begin{widetext}
\begin{eqnarray}
j_{e} ={2e W\over \pi ^2 } \int_0^{eV/\hbar v_F}dk_x   \int _{\sqrt{(eV/\hbar v_F)^2-k_x^2}}^\infty dk_y  v_x T
={2ev_F W\over \pi^2} \int_0^{eV/\hbar v_F} dk\int _{k}^\infty dk_y \frac {p^2}{p^2 +k_y^2\sinh^2(pL) }
\nonumber \\
={ev_F W\over \pi^2}\int _0^\infty {p\,dp \over \cosh(pL)} \ln{\sqrt{(\hbar v_Fp/e)^2+V^2} \cosh(pL)+V\over \sqrt{(\hbar v_Fp/e)^2+V^2} \cosh(pL)-V},
           \end{eqnarray}
\begin{eqnarray}
{dj_e\over dV} = G_0L \int_0^\infty {p^3\,dp\over \sqrt {(eV/\hbar v_F)^2+p^2} 
[(eV/\hbar v_F)^2 \sinh^2(pL) +p^2 \cosh^2(pL) ]}.
     \end{eqnarray}
           \end{widetext}
At high voltages  $V\to \infty$ this current saturates at the value 
\begin{eqnarray}
j_{es} ={2ev_F W\over \pi^2}\int _0^\infty {p\,dp \over \cosh(pL)} \ln \coth{pL\over 2} \approx  {1.82 ev_F W\over \pi^2 L^2}.
           \end{eqnarray}

\begin{figure}%[t]
\centerline{\includegraphics[width=0.8\linewidth]{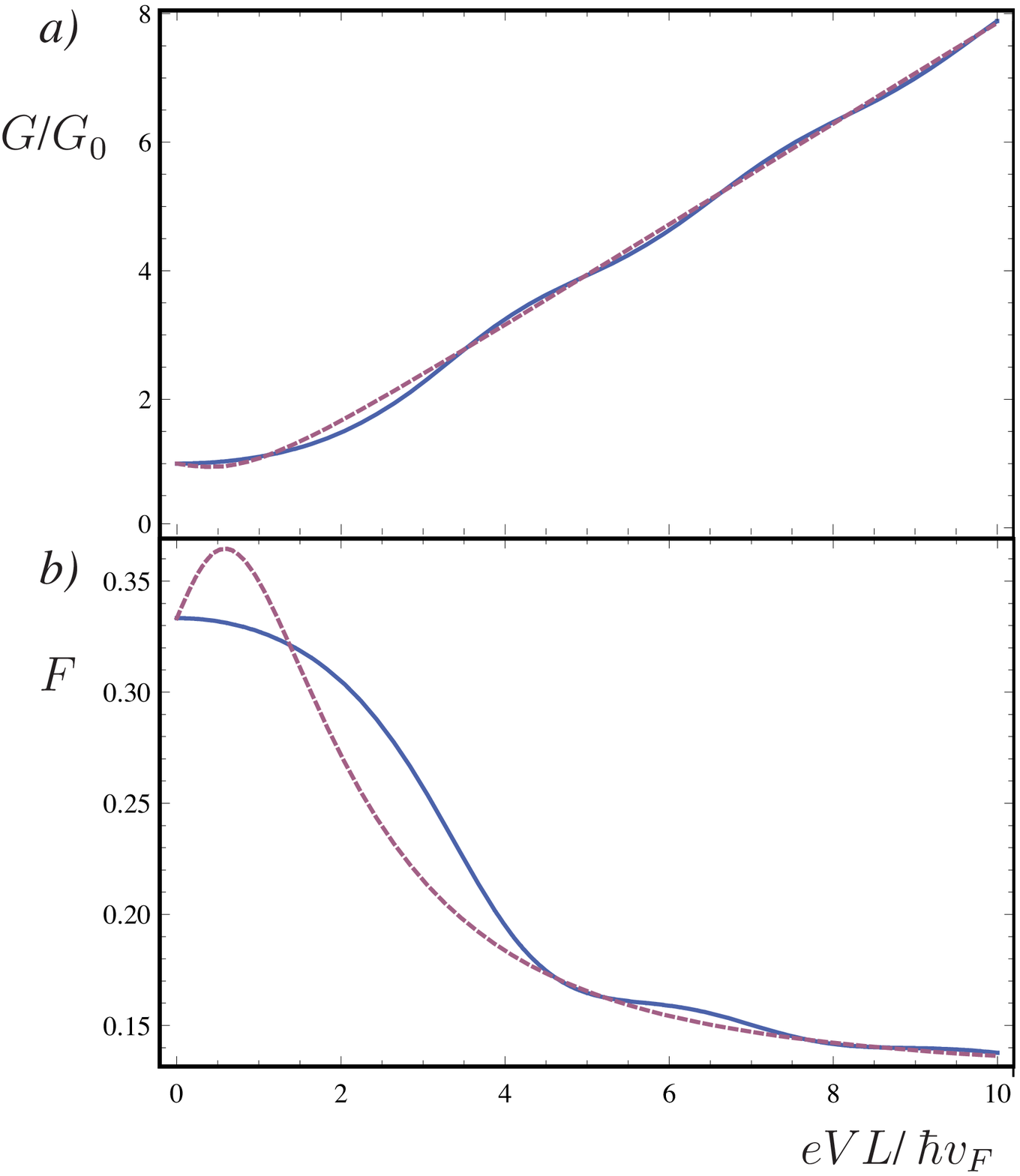}}
\caption{(color online) The relative conductance $G /G_0$ (a) and the Fano factor $F$ (b) as functions of $V=V_1$ ($V_2=0$). Solid lines refer to the phase-coherent propagation through the sheet, and  dashed lines refer to  the phase-incoherent case.} 
\label{fig2}
\end{figure}

Now let us write down the expressions for shot noise power. The contribution from evanescent modes:
\begin{eqnarray}
S_{e} ={4e^2 W\over \pi^2  } \int_0^{eV/\hbar v_F}dk_x   \int _{\sqrt{(eV/\hbar v_F)^2-k_x^2}}^\infty dk_y  v_x T(1-T)
%\nonumber \\
={4e^2v_F W\over \pi^2} \int_0^{eV/\hbar v_F} dk\int _{k}^\infty dk_y \frac {p^2k_y^2\sinh^2(pL) }{[p^2 +k_y^2\sinh^2(pL) ]^2}.
           \end{eqnarray}
The propagating modes contribute the power 
\begin{eqnarray}
S_p ={4e^2 W\over \pi^2  } \int_0^{eV/\hbar v_F}dk_x   \int _0^{\sqrt{(eV/\hbar v_F)^2-k_x^2}} dk_y  v_x T(1-T)
%\nonumber \\
={4e^2v_F W\over \pi^2} \int_0^{eV/\hbar v_F} dk\int _0^{k} dk_y \frac {k_x^2k_y^2\sin^2(k_xL)}{[k_x^2 +k_y^2\sin^2(k_xL) ]^2}.
      \label{shot}   
       \end{eqnarray}
At high voltages $V\to \infty $ 
\begin{eqnarray}
S_p ={0.196 ev_F W\over \pi^2 }\left({eV\over \hbar v_F}\right)^2.
       \end{eqnarray}
Solid lines in Fig. \ref{fig2} show the total conductance $G =d(j_p+j_e)/dV$ and the Fano factor $F=(S_p+S_e)/2e(j_p+j_e)$ as functions of $V=V_1$. The Fano factor changes from 1/3 at low voltages to 0.125 at high voltages. 

As well as in  Refs. \onlinecite{Kats} and  \onlinecite{Been}, we assumed up to now that a particle propagated along the sheet  coherently without any dephasing.
Meanwhile, even in the ballistic regime the electron propagating across the sheet not necessarily ``remembers'' its phase. It is known\cite{Dat} that in the absence of the phase coherence the total transmission through the double contact is combined not from amplitudes of transmission but from probabilities of transmission. Namely, for the incoherent case one should replace Eq. (\ref{cor}) by  
\begin{eqnarray}
T=1-R={T_1T_2\over 1-R_1R_2}=\frac {k_x}{k },
           \end{eqnarray}
where $T_{1,2}=|t_{1,2}|^2$ and $R_{1,2}=|r_{1,2}|^2$. Then the expressions Eqs. (\ref{prop}) and (\ref{shot}) for the current and the shot-noise power should be replaced 
by
\begin{eqnarray}
j_p ={2ev_F W\over \pi^2} \int_0^{eV/\hbar v_F} dk\int _0^{k} dk_y \frac {k_x}{k}
%\nonumber  \\
={0.785 ev_F W\over \pi^2}\left({eV\over \hbar v_F}\right)^2,
      \label{propg}     \end{eqnarray}
\begin{eqnarray}
S_p={4e^2 v_F W\over \pi^2} \int_0^{eV/\hbar v_F} dk\int _0^{k} dk_y \frac {k_x(k-k_x)}{k^2}
%\nonumber \\
={0.22 e^2v_F W\over \pi^2}\left({eV\over \hbar v_F}\right)^2.
%      \label{prop}    
       \end{eqnarray}
The conductivity and the Fano factor for the incoherent propagation are shown in Fig. \ref{fig2} with dashed lines. The comparison of two cases shows that at high voltages  the role of possible phase coherence is not essential, and the oscillations resulting from the phase dependence are insignificant as was observed in the experiments\cite{Marc,Hak}. Apparently the reason of it is that even in the phase-correlated case there is decoherence from summation over the states with different wave vectors. The most remarkable difference between the two cases  is non-monotonicity  of   dependences in the incoherent case at low voltages: the conductiance has a minimum, where the conductance is slightly less than the conductance at $V=0$, (so the latter is not the minimum conductance anymore!), whereas the Fano factor has a maximum, where its value 0.37  exceeds 1/3.  It is worthwhile to note that DiCarlo {\em et al.}\cite{Marc} observed the Fano factor up to 0.38, which could be an evidence of incoherent propagation in the sheet. 

A similar contribution can be done for  nonzero voltage $V=V_2$ (with $V_1=0$)  independently on whether the charge is transported only by electrons ($V_2>0$) or also by holes ($V_2<0$). Symmetry between electron and holes is valid as far as the Fermi wave vectors in the leads essentially exceed the wave vectors $k$ inside the sheet.  In general the total current through the sheet and the Fano factor are determined by the relations
\begin{eqnarray}
J  =j_p(V_1) +j_e(V_1) -j_p(V_2) -j_e(V_2)
      %\label{prop}
           \end{eqnarray}
and
\begin{eqnarray}
F  ={1\over 2e}\frac{S_p(V_1) +S_e(V_1) -S_p(V_2) -S_e(V_2)}
{j_p(V_1) +j_e(V_1) -j_p(V_2)-j_e(V_2)}.
      %\label{prop}
           \end{eqnarray}

One might expect that the ballistic transport itself should be lossless and noiseless. But it is well known that transport through any ballistic conductor is inevitably accompanied by dissipation and noise at the contacts\cite{Dat}: electrons, which are injected from one lead to another, cross the ballistic conductor without scattering,  but disturb the equilibrium in the lead, where they are injected. This results in dissipation inside leads. Resistance and noise for the transport  through graphene also originate from stochastic processes of relaxation in the leads near the contacts. But the unique feature of graphene is that the contact phenomena are governed by  the peculiar spectrum in the Dirac point rather than by details of the spectrum in leads. 

In summary, we analyzed the current and the Fano factor as functions of voltage in a graphene sheet for arbitrary voltage drops between leads and the sheet in the limit of infinite aspect ratio of the sheet width  $W$ to its length $L$. The oscillations on these dependences, which were revealed earlier\cite{Been} for the finite aspect ratio, are insignificant in the limit $W/L\to \infty$. The case of incoherent ballistic transport was also analyzed. At high voltages  the difference with coherent transport is not essential. But at low voltages conductance and and Fano-factor dependences for incoherent transport become non-monotonous so that the conductance has a minimum and the Fano factor has a maximum at non-zero voltage bias.

I thank Pertti Hakonen and Yakov Kopelevich  for discussions, which stimulated this work. The work was supported by the Large Scale Installation Program ULTI-3 of the European Union.


\begin{thebibliography}{99}
\bibitem{Kop} I. A. Luk'yanchuk and Y.  Kopelevich, Phys. Rev. Lett. {\bf 93}, 166402 (2004).

\bibitem{Geim} K. S. Novoselov, A. K. Geim, S. V. Morozov, D. Jiang,
M. I. Katsnelson, I. V. Grigorieva, S. V. Dubonos, and A. A. Firsov, Nature {\bf 438}, 197 (2005).

\bibitem{Kim}  Y. Zhang, Y.-W. Tan, H. L. Stormer, and P. Kim,
Nature {\bf 438}, 201 (2005).
\bibitem{Lau} F. Miao, S. Wijerante, Y. Zhang, U. C. Coskun, W. Bao, and C. N. Lau, Science {\bf 317}, 1530 (2007).
\bibitem{Kats} M. I. Katsnelson, Eur. Phys. J. B {\bf 51}, 157 (2006).
\bibitem{Been} J. Tworzyd{\l}o, B. Trauzettel, M. Titov, A. Rycerz, and C. W. J. Beenakker,
  Phys. Rev. Lett. {\bf 96}, 246802 (2006).
  \bibitem{Marc} L. DiCarlo, J. R. Williams, Y. Zhang, D. T. McClure, and C. M. Marcus, Phys. Rev. Lett. {\bf 100}, 156801 (2008).
  \bibitem{Hak} R. Danneau, F. Wu, M. F. Craciun, S. Russo, M. Y. Tomi, J. Salmielto, A. F. Morpugo, and P. J. Hakonen, archiv: cond-mat/0711.4306.

\bibitem{schom}  H. Schomerus, Phys. Rev. B {\bf 76}, 045433 (2007).
\bibitem{blant}   Ya. M. Blanter and I. Martin,   Phys. Rev. B {\bf 76}, 155433 (2007).
\bibitem{Dat} S. Datta, {\sl Electronic Transport in Mesoscopic Systems} (Cambridge 
University Press, Cambridge, 1997).
  
  
  
\end{thebibliography}
\end{document}